\documentclass[twocolumn,showpacs,prb,amsmath,amssymb]{revtex4}
\usepackage{graphicx}% Include figure files
\usepackage{dcolumn}% Align table columns on decimal point
\usepackage{bm}% bold math
\usepackage{hyperref}
\DeclareGraphicsExtensions{.eps}
\graphicspath{{figures/}}
\usepackage{subfigure}
\begin{document}

%opening
\title{Exchange interactions and magnetic phases of transition metal oxides: benchmarking  advanced {\em ab initio} 
methods.}
\author{T. Archer, C.D. Pemmaraju and S. Sanvito}
\affiliation{School of Physics and CRANN, Trinity College Dublin, Ireland}
\author{C. Franchini and J. He}
\affiliation{University of Vienna, Faculty of Physics and Center for Computational Materials Science, A-1090 Vienna, 
Austria}
\author{A. Filippetti, P. Delugas, D. Puggioni, and V. Fiorentini} 
\affiliation{CNR-IOM, UOS Cagliari and Dipartimento di Fisica, Universit\`a di Cagliari, Monserrato (CA), Italy}
\author{R. Tiwari and P. Majumdar} 
\affiliation{Harish-Chandra Research Institute, Chhatnag Road, Jhusi, Allahabad 211019, India}
\date{\today}

%%%%%%%%%%%%%%%%%%%%%%%%%%%%%%%%%%%%%%%%%%%%%%%%%%%%%%%%%%%%%%%%%%%%%%%%%%%%%%%%%%%%%%%%%%%%%%%%%%%%%%%%%%%%%%%%%%%%%%
%									ABSTRACT	
%%%%%%%%%%%%%%%%%%%%%%%%%%%%%%%%%%%%%%%%%%%%%%%%%%%%%%%%%%%%%%%%%%%%%%%%%%%%%%%%%%%%%%%%%%%%%%%%%%%%%%%%%%%%%%%%%%%%%

\begin{abstract}
The magnetic properties of the transition metal monoxides MnO and NiO are investigated at equilibrium and under pressure
via several advanced first-principles methods coupled with  Heisenberg Hamiltonian MonteCarlo. The comparative 
first-principles analysis involves two promising beyond-local density functionals approaches, namely the hybrid density 
functional theory and the recently developed variational pseudo-self-interaction correction method, implemented with both 
plane-wave  and atomic-orbital basis sets. The advanced functionals deliver a very satisfying rendition, curing the main 
drawbacks of the local functionals and improving over many other previous theoretical predictions. Furthermore, and most 
importantly, they convincingly demonstrate a degree of internal consistency, despite differences emerging due to 
methodological details (e.g. plane waves vs. atomic orbitals). 
\end{abstract}


% insert suggested keywords - APS authors don't need to do this
%\keywords{Self-interaction correction, hybrid functionals, DFT, transition metal monooxides, exchange coupling, 
%N\'eel temperature, Montecarlo}

%\maketitle must follow title, authors, abstract, \pacs, and \keywords
\maketitle

\section{Introduction}

The relative simplicity of structural and magnetic ordering and the abundance of available experimental and theoretical 
data elect transition metal monoxides (TMO) as favorite prototype materials for the {\em ab initio} study of exchange 
interactions in Mott-like insulating oxides.\cite{anderson59} TMO are known to be robust antiferromagnetic (AF) Mott 
insulators with sizable exchange interactions and N\'eel ordering temperatures. The accurate determination of magnetic 
interactions purely by first-principles means is a remarkable and as yet unsolved challenge.\cite{solovyev,wan,kotani} 
The difficulty stems, on the one hand, from fundamental  issues in the description of Mott insulators by standard density
functional theory (DFT) approaches, such as local-spin density approximation (LDA) or the generalized gradient 
approximation (GGA). On the other hand, the determination of low-energy spin excitations require a meV-scale accuracy; 
however, the error bar  due to specific implementation and technical differences may easily be larger.

A large amount of theoretical work for TMO has amassed over the years. A number of studies were carried out in particular
for MnO and NiO with a variety of advanced methods: the LDA+U,\cite{dudarev, fang, solovyev, rohrbach, pask01} 
GGA+U,\cite{zhang,bayer07} the optimized effective potential (OEP),\cite{solovyev} the quasiparticle Green function (GW)
approach,\cite{kotani,arya,massidda} several types of self-interaction corrected LDA (SIC-LDA),
\cite{svane,szotek,kodde,fischer,dane,huges} Hartree-Fock,\cite{towler, bredow, franchini, moreira} and several types of hybrid functionals such as 
B3LYP,\cite{feng, moreira} PBE0,\cite{franchini,tran,chen}, Fock-35\cite{moreira}, and B3PW91.\cite{tran} From a methodological viewpoint, 
Refs. \onlinecite{kasinathan}, \onlinecite{solovyev}, and \onlinecite{fischer} are particularly relevant for our present
purposes, because of the systematic  comparison of  diverse approaches  to computing magnetic interactions. 
Other studies\cite{cohen, mattila, kunes} focused in particular on pressure-induced high-to-low spin magnetic collapse 
observed at very high pressure ($\sim$150 GPa for MnO) and relevant to Earth-core geophysics. Here we will not, however,
be concerned with the phenomenology of this specific phase transition.

The present work presents a detailed theoretical analysis of MnO and NiO magnetic properties on a wide range of lattice 
parameters (i.e. hydrostatic pressures) carried out by an array of both standard and advanced first-principles 
methods. In particular, our theoretical front-liners are two approaches proposed recently for the description of 
strongly-correlated systems: the Heyd, Scuseria and Ernzerhof (HSE) hybrid functional approach,\cite{hse} and the recently developed variational pseudo
self-interaction correction,\cite{vpsic} implemented in two different methodological frameworks: plane-wave basis set 
plus ultrasoft\cite{uspp} pseudopotentials (PSIC), and linear combination of atomic-orbital basis set (ASIC). To provide 
a baseline for their evaluation, we complement these methods by their local counterparts implemented in the same 
methodological setting, namely LDA  in plane-waves and ultrasoft pseudopotentials (reference for PSIC), local-orbitals and
norm-conserving pseudopotentials (reference for ASIC), and GGA in the Perdew-Becke-Ernzerhof\cite{pbe}(PBE) version 
(reference for HSE). Performing the same set of calculations in parallel with  different methods, as well as different  
implementations of the same method is instrumental to distinguish fundamental and methodological issues, and characterizes
this work with respect to the many previous theoretical studies of TMO.

MnO and NiO in equilibrium conditions have a high-spin magnetic configuration and large ($\sim$ 3.5-4 eV) band gap. 
Magnetic moments and exchange interactions depend crucially on the details of structural and electronic properties. 
The latter are characterized by a complex interplay of distinct energy scales: the crystal field splitting, which in 
rocksalt symmetry separates the on-site 3d $e_g$ and $t_{2g}$ energies; the charge transfer energy between O $p$ and 
TM d states; the hopping energies between d-d and p-d states.\cite{zaanen} Ref.\onlinecite{solovyev} convincingly shows 
that an empirical single-particle potential suitably adjusted to reproduce the experimental values of the above mentioned
energies can deliver highly accurate magnetic interactions (moments and spin-wave dispersion). However, obtaining a correct
balance of all these contributions is difficult even for advanced density functional methods, not to mention standard LDA 
or GGA, which fails altogether (to different extents depending on the specific compound). A general analysis of these 
failures and difficulties of {\em ab initio} approaches is beyond the scope of this work; we mention that the thorough 
analysis carried out in Ref.\onlinecite{solovyev} and \onlinecite{nekrasov08} suggests that a single parameter, as adopted
by the LDA+U, is not sufficient, while global multi-state energy corrections could serve this purpose. Both the advanced 
methods (HSE and PSIC/ASIC) adopted in this work, while quite different in spirit, act in terms of "global" corrections 
to local density functional energy spectrum, i.e. no a priori assumption is made about which particular state or band is 
in need of modification or correction.

A large portion of the LDA/GGA failure in the description of correlated systems can be imputed to the presence of 
self-interaction (SI) -- that is, the interaction of each single electron with the potential generated by itself. 
One consequence of SI is a severe, artificial upshift in energy of the on-site single-particle energies in the 
density-functional Kohn-Sham  equations (compared e.g.  to measured photoemission energies). Spatially localized states 
are naturally characterized by a large SI: these will include 3d states, but also (albeit to a lesser extent) 2p states, 
and of course localized interface, surface, and point-defect states. It is sometime argued that this error is immaterial 
because Kohn-Sham eigenvalues are not supposed to match single-particle energies, and density-functional theories
are only valid for ground-state properties.
%However, highest-occupied and lowest-unoccupied electronic states are exceptions, since they must be rigorously related to emission and adsorption energies, and as such related to ground-state properties.
However, highest-occupied and lowest-unoccupied electronic states are exceptions,
since the former must be rigorously related to the first ionization energy and thus is a ground-state property while the latter may also be interpreted within generalized KS schemes [Phys Rev B 53, 3764, (1996)] as an approximate electron addition energy.
As a consequence, the SI error may dramatically affect the predicted ground state properties, even turning 
the Kohn-Sham band structure from insulating to metallic (a frequent occurrence for Mott-like systems in LDA/GGA).

MnO and NiO are both affected by SI, although to different extents:\cite{ff-coll} severely for MnO, dramatically for 
NiO. Under positive (i.e. compressive) pressure the problem will be  amplified, as any small band gap which may exist at 
equilibrium will be further reduced up to complete closure, and magnetic moments may be disrupted. Thus, a reliable 
description of TMO under pressure necessarily requires approaches overcoming the SI problem. Both PSIC/ASIC and HSE, 
although from different starting points, work towards the suppression of SI. The former explicitly subtracts the SI 
(written in atomic-like form) from the LDA functional; the latter, in a more fundamental manner, inserts of a portion of 
true Fock exchange in place of the local exchange functional, whose incomplete cancellation with the diagonal Hartree 
counterpart is the source of SI in LDA/GGA functionals.\cite{becke} Our results will show that, despite the different 
conceptual origin, the two approaches deliver a systematically consistent description of MnO and NiO, in fact with 
spectacular quantitative agreement in several instances.

The paper is organized as following: Section \ref{met} describes briefly  the  methodologies employed; in 
Section \ref{magn_struc}, the model used to calculated the exchange-interaction parameters is discussed. 
Section \ref{res}  illustrates our results at equilibrium (\ref{res_equil}) and under pressure (\ref{res_press}). 
In Sections \ref{res_exch} and \ref{res_crit} we discuss the exchange interactions and critical temperatures, 
respectively. Section \ref{concl} offers some concluding remarks. 

\section{Methods}
\label{met}

Our first-principles results are obtained by using  three different codes: PWSIC\cite{pwsic}, SIESTA\cite{siesta} and VASP\cite{vasp}.
With the PWSIC code, which uses plane-waves basis set and ultrasoft pseudopotentials we carry out calculations within LDA 
(hereafter  LDA-PW) and PSIC. The SIESTA\cite{siesta} package, implemented over a local (atomic) orbital basis set and norm-conserving 
pseudopotentials, was employed for LDA (LDA-LO below) and ASIC calculations (note that the ASIC is available only in a developer version of the
code). The variational PSIC/ASIC\cite{vpsic} approach evolves from the non-variational precursors PSIC\cite{fs,ff-coll} and ASIC.\cite{asic} The new 
method describes electronic properties as accurately as its precursors, and variationality allows its application to structural optimization and 
frozen-phonon calculations via quantum forces. In the following we briefly sketches the main features of the method, remanding the reader to 
Ref.~[\onlinecite{vpsic}] for the detailed formulation for both plane waves (PSIC) and local orbital (ASIC) basis set. 

The PSIC starts from the following energy functional
\begin{equation}
E^{PSIC}[\{\psi\}]\,=\,E^{LDA}[\{\psi\}]-{1\over 2}\sum_{ij\nu\sigma}\,\epsilon^{SI}_{ij\sigma\nu}[\{\psi\}]\,
p^{\sigma}_{ji\nu}[\{\psi\}]
\label{vpsic_etot}
\end{equation}
where here $i$, $j$ are collective quantum numbers ($l_i$, $m_i$, $l_j$=$l_i$, $m_j$) of a minimal basis (i.e. one for each angular moment $l$) 
of atomic wavefunctions, and $\sigma$ and $\nu$ are the spin and the atomic site index, respectively. Thus, to the usual LDA energy a SI contribution is 
subtracted out. This is written as single-particle SI energies ($\epsilon^{SI}$) scaled by the effective orbital occupations ($p$). For an extended system, whose 
eigenfunctions are Bloch states ($\psi_{n{\bf k}}^{\sigma}$), the orbital occupations are self-consistently calculated as projection of the Bloch states 
onto the atomic orbitals $\phi$
\begin{eqnarray}
P^{\sigma}_{ij\nu}=\,\sum_{n{\bf k}}\, f_{n{\bf k}}^{\sigma}\, \langle\psi_{n{\bf k}}^{\sigma}|\phi_{i,\nu}\rangle \> \langle \phi_{j,\nu} 
|\psi_{n{\bf k}}^{\sigma}\rangle,
\label{occ}
\end{eqnarray}

In addition, the ``effective'' SI energies are defined as 
\begin{eqnarray}
\epsilon^{SI}_{ij\sigma\nu}\,=\,\sum_{n{\bf k}}\, f_{n{\bf k}}^{\sigma}\, \langle\psi_{n{\bf k}}^{\sigma}|
\gamma_{i,\nu}\rangle C_{ij} \langle\,\gamma_{j,\nu}|\psi_{n{\bf k}}^{\sigma}\rangle
\label{epsi-si}
\end{eqnarray}
where $\gamma_{i,\nu}$ is the projection function of the SI potential for the $i^{th}$ atomic orbital centered
on atom $\nu$. In the radial approximation one has
\begin{eqnarray}
\gamma_{l_i,m_i,\nu}({\bf r})\>=\> V_{HXC}[\rho_{\nu,l_i}(r);1]\>\phi_{l_i,m_i,\nu}({\bf r})\:,
\label{gamma}
\end{eqnarray}
with $C_{ij}$ being normalization coefficients
\begin{eqnarray}
C^{-1}_{l_i,m_i,m_j}=\int d{\bf r}\phi_{l_i,m_i}({\bf r})\,V_{HXC}[\rho_{\nu,l_i}(r);1]\,\phi_{l_i,m_j}({\bf r})\:.
\label{cij}
\end{eqnarray}
It is now easy to see that the functional derivative of this PSIC energy functional leads to Kohn-Sham single particle Equations similar to those
adopted in the original scheme described in Ref.~[\onlinecite{fs}].

The VASP\cite{vasp} code, emloying the projected augmented wave (PAW) approach, is used for PBE\cite{pbe} and 
HSE\cite{hse} calculations. In the HSE method the short-range (sr) part of the exchange interaction (X) is constructed 
by a proper mixing of exact non-local Hartree-Fock exchange and approximated semi-local PBE exchange.
The remaining contributions to the exchange-correlation energy, namely the long-range (lr) exchange interaction and 
the electronic correlation (C), is treated at PBE level only, resulting in the following expression:
\begin{equation}
E_{XC}^{HSE} = \frac{1}{4}E_X^{HF,sr,\mu} + \frac{3}{4}E_X^{PBE,sr,\mu} + E_X^{PBE,lr,\mu} + E_C^{PBE}\:.
\end{equation}
The partitioning between sr and lr interactions is achieved by properly decomposing the Coulomb kernel (1/$r$, 
with $r= | {\bf r}-{\bf r'} |$) through the characteristic parameter $\mu$, which controls the range separation between 
the short (S)  and long (L) range part
\begin{equation}
\frac{1}{r}=S_{\mu}(r) + L_{\mu}(r)=\frac{\rm erfc(\mu{r})}{r} + \frac{\rm erf(\mu{r})}{r}\:.
\end{equation}
We have used here $\mu=0.20$ $\AA^{-1}$, in accordance to the HSE06 parameterization\cite{hse06} and corresponding to the
distance 2/$\mu$ at which the short-range interactions become negligible. For $\mu$= 0, HSE06 reduces to the 
parent unscreened hybrid functional PBE0\cite{pbe0}.
In terms of the one-electron Bloch states {$\phi_{{\bf k}n}({\bf r})$} and the corresponding occupancies {$f_{{\bf k}n}$}
the sr Hartee-Fock exchange energy $E_X^{HF,sr,\mu}$ can be written as
\begin{eqnarray}
E_X^{HF,sr,\mu} & =     & -\frac{e^2}{2} \sum_{{{\bf k}n},{{\bf q}m}} 2w_{\bf k}f_{{\bf k}n} \times w_{\bf q}f_{{\bf q}m} \\
                &\times & \int \int d^3{\bf r} d^3{\bf r}'\frac{{\rm erfc}(\mu|{\bf r}-{\bf r}'|)}{|{\bf r}-{\bf r}'|}\\
                &\times & \phi_{{\bf k}n}^{*}({\bf r})\phi_{{\bf q}m}({\bf r})\phi_{{\bf q}m}^{*}({\bf r}')\phi_{{\bf k}n}({\bf r}')\:,
\end{eqnarray}
from which one can then derive the corresponding non-local sr Hartee-Fock exchange potential\cite{marsman}.
In the last few years the application of the HSE06 to a wide class of solid state systems, including archetypical
benchmark examples\cite{paier}, transition metal oxides\cite{franchini, franch07, franchbab}, dilute magnetic 
semiconductors\cite{droghetti,stroppa11}, multifferoics\cite{stroppa10} and magnetic perovskites\cite{ustc}, has demonstrated,
that HSE06 delivers a substantially improved description of ground state structural, electronic, magnetic and vibrational properties
with respect to the standard PBE and PBE+U. However, the use of hybrid functionals for metals in less satisfactory, 
because of the general significant overestimated bandwidth\cite{paier}.

As far as the technical aspects are concerned, PWSIC calculations have been carried out by using 16-atom face centered cubic (FCC) 
supercells [i.e. 8 formula units (f.u.)], cut-off energies of 40 Ry, reciprocal space integration over 6$\times$6$\times$6 and 10$\times$10$\times$10 
special k-point grids for self-consistency and density of states calculations, respectively. VASP calculations have been performed using a 4 f.u. unit 
cell, an energy cut-off of 25 Ry, a 4$\times$8$\times$4 k-point mesh and a standard HSE mixing parameter $a$=0.25. The SIESTA calculation 
have been performed using a 4 f.u. cell with a 6$\times$11$\times$6 k-point mesh, a real space mesh cut-off of 800 Ry and a double-$\zeta$ 
polorized basis. Pressures have been evaluated using the  Birch-Murnaghan equation of state.\cite{birch}

Montecarlo simulations of the classical 2-parameter Heisenberg model  have been carried out for a spin lattice system of size L=12 (i.e. N = L$^3$ total 
lattice sites). We determined ground state magnetic ordering and critical temperature by simulated annealing for each pair of ab-initio-calculated 
J$_1$ and J$_2$ parameters characterizing the magnetic structure (see next Section), at each lattice constant and for each method. In order to tests 
finite-size effects on the results some annealing with L=20 was also performed. The annealing  was done over 30 temperature points, starting from high 
temperature (roughly twice the critical temperature as estimated by few trial runs) down to T=0, with 10$^6$ sweeps at each temperature. The annealing 
protocol is the usual Metropolis algorithm based on single spin update.
 
\section{Magnetic structures and the Heisenberg model}
\label{magn_struc}

\begin{figure}[ht]
\centerline{\includegraphics[clip,width=8.5cm]{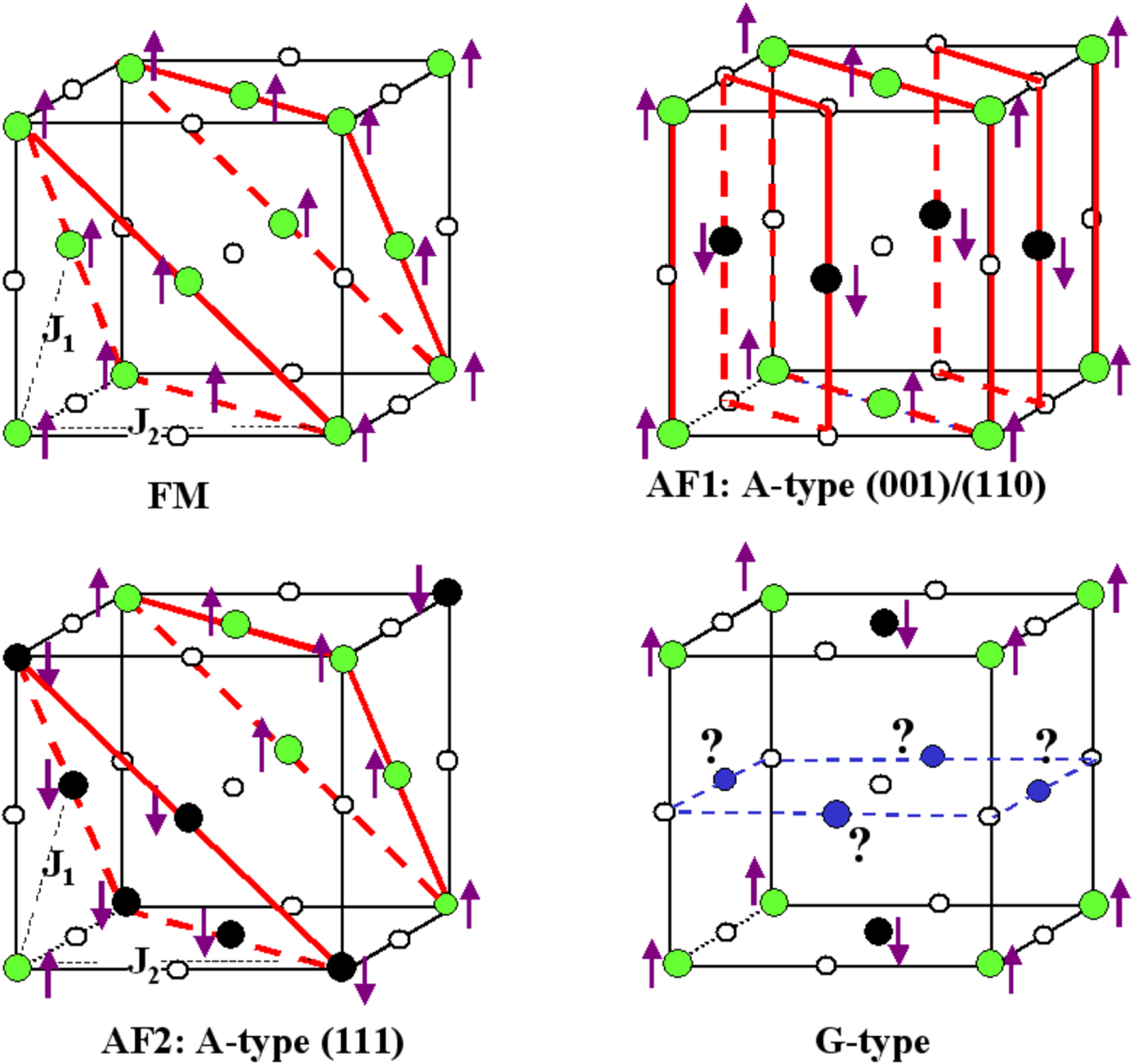}}
\caption{(Color on-line) Magnetic phases used for the Heisenberg model fit: FM; AF$_2$ that is built of alternating (111) planes of like spins
[highlighted by thick (red) lines]; AF$_1$, made of alternating (011) (or equivalently (001)) planes of like spins,  delimited by (red) thick lines. 
The highly frustrated G-type AF phase is also shown for comparison. Filled light (green, up spin) and black (down spin) circles
indicate TM atoms, small empty circles are for oxygens. For the G-type phase (blue) filled circles marked with a question mark indicate TM atoms with 
frustrated spin direction.
\label{sketch}}
\end{figure}

TMO have a rock-salt structure (see Fig.\ref{sketch}), so each TM has 12 nearest-neighbor (NN) and 6 next-nearest neighbors (NNN). The NNN are connected  
through  oxygen bridges, and  their interaction J$_2$ is dominated by superexchange. On the other hand, NN interact via a typically smaller exchange 
coupling J$_1$ whose sign may depend on the specific TMO; J$_{1}$ involves direct TM-TM exchange (giving a robust AF contribution) and a  
90$^{\circ}$-oriented TM-O-TM superexchange (expected to be weakly FM). The observed ground state magnetic phase is antiferromagnetic (111) A-type, 
labeled AF$_2$ hereafter. It can be seen as a stacking of (111) planes of like spin alternating along the [111] direction, as illustrated in 
Fig. \ref{sketch}. In AF$_2$ each TM has 6 spin-paired intra-(111)-plane NN and 6 spin-antipaired inter-(111)-plane NN; on the other hand, all 6 NNN bonds
are inter-planar and antipaired.  Thus, this configuration maximizes the energy gain associated to the NNN antiparallel spin alignment. As for beyond-NNN
magnetic interactions, there is ample experimental\cite{hutchings} and theoretical\cite{fischer} evidence that they can be safely discarded
(e.g. according to inelastic neutron scattering\cite{hutchings} in NiO they are two order of magnitude smaller than the dominant J$_2$. We explicitly 
checked this with the SIESTA code.).   

In order to evaluate J$_1$ and J$_2$ we need to consider at least two competing high-symmetry magnetic phases  beside the observed AF$_2$. Natural choices
are the ferromagnetic (FM) order and the AF (110) A-type order with (110) spin-paired planes compensated along [110] (labeled AF$_1$). AF$_{1}$  can also 
be seen as made of FM (001) planes alternating along [001] (see Fig. \ref{sketch}). The AF$_1$ phase has all the 6 NNN spin-paired, 4 of the NN 
spin-paired, and 8 NN spin-antipaired. On the other hand, it is interesting to note that the G-type AF order (also depicted in Fig. \ref{sketch}) is 
strongly disfavored by frustration, since in FCC symmetry there is no way to arrange the 12 NN interactions in antiparallel fashion without
conflict. 

To extract J$_1$ and J$_2$ we fit our calculated total energies to a standard 2-parameter classical Heisenberg Hamiltonian 
of the form:

\begin{eqnarray}
H=- J_1 \sum_{\langle i,j\rangle} \vec{e_i}\cdot\vec{e_j} -J_2 \sum_{\langle\langle i,j\rangle\rangle} \vec{e_i}\cdot\vec{e_j}
\label{heisen}
\end{eqnarray}

where $\langle i,j\rangle$ and $\langle\langle i,j\rangle\rangle$ indicate summation over NN and NNN, respectively, and $\vec{e_i}$ is the spin direction
unit vector. Energies (per f.u.) are then expressed as:

\begin{eqnarray}
E_{FM} & = & E_0  - 6J_1 -3J_2 \nonumber\\
E_{AF_1} &= & E_0  + 2J_1 -3J_2 \nonumber \\
E_{AF_2} &= & E_0 +3J_2
\end{eqnarray}

This is solved to give:
\begin{eqnarray}
J_1 & = \frac{1}{8} (E_{AF_1}-E_{FM})  \nonumber\\
J_2 & = \frac{1}{24} (4E_{AF_2}-3E_{AF_1}-E_{FM}) \label{EqJ}
\end{eqnarray}

With this choice of the Hamiltonian, negative and positive J values correspond to energy gain for spin-antiparallel and spin-parallel orientations, 
respectively.

Finally, we mention that several other anisotropic terms may in principle contribute to the Heisenberg Hamiltonian, 
related to short-range dipolar interactions favoring a preferential spin direction parallel to (111) planes,
and to rhombohedral distortions of the AF$_2$ phase consisting on a (111) inter-planar contraction and 
slight change of the perfect 90$^{\circ}$ angle of the rock-salt cell, which causes a symmetry breaking of J$_1$ in two 
J$_1^+$ and J$_1^-$ values.\cite{franchini} However, all these effects are quantified to be order-of-magnitude smaller than the dominant
exchange-interaction energies (e.g. for NiO J$_1^+$-J$_1^-$ $\sim$ 0.03 meV according to neutron data\cite{hutchings}),
thus the Heisenberg Hamiltonian written in Eq.\ref{heisen} can be considered fully sufficient for our present purposes.

\section{Results}
\label{res}

\subsection{Equilibrium structures}
\label{res_equil}

We have calculated total energies and pressures of MnO and NiO as a function of lattice parameter for the 3 magnetic 
phases FM, AF$_2$, and AF$_1$. Values of the equilibrium lattice parameter and bulk modulus for the stable 
AF$_2$ phase are reported in Table \ref{tab_lattice}, in comparison with the experimental values. 

\begin{center}
\begin{table}
\caption{Equilibrium lattice constants a$_0$ (in \AA) and bulk moduli B$_0$ (in GPa) calculated in this work with 
various methods, compared with experimental values. All values refer to the stable AF$_2$ magnetic ordering.} 
\begin{tabular}{lccccccc}
\hline 
         & LDA-PW       & PSIC            & LDA-LO    & ASIC        & PBE   & HSE &   Expt. \\ 
\hline\hline  
\multicolumn{8}{c}{MnO} \\
%MnO         &            &                 &          &          &       &      &           \\  
a$_0$  & 4.38       &  4.35                &   4.35         &   4.35   & 4.47  & 4.41 &   4.43$^a$ \\
B$_0$  &  158       &  194                &   177    &  199  & 145   &  170 &   151$^b$,162$^c$ \\
\hline\hline
\multicolumn{8}{c}{NiO} \\
%NiO         &            &                 &            &          &       &       &          \\  
a$_0$  & 4.15       &  4.09           &    4.09    &   4.06   & 4.19  & 4.18        &   4.17$^d$   \\
B$_0$  &  234       &  269            & 270   & 287    &  183  &  202        &  180-220$^e$  \\
\hline\hline 
 \multicolumn{8}{l} { a): Ref.\onlinecite{sasaki};  b): Ref.\onlinecite{noguchi}, c): Ref.\onlinecite{jeanloz} } \\
 \multicolumn{8}{l} { d): Ref.\onlinecite{schmahl}; e): Ref.\onlinecite{huang}   } \\   
\end{tabular}
\label{tab_lattice}
\end{table}
\end{center}

Results are quite satisfactory overall: each method predicts an equilibrium lattice constant in good agreement (within 1-2\%) with 
experiment for the AF$_2$ phase. It is well known that structural properties calculated by LDA or GGA can be good, 
or even excellent, although the electronic properties are poor.\cite{marsman,franch07} The results for MnO and NiO are 
a case in point, as both LDA-PW and PBE stay within 1{\%} from the experiment (the former in defect, the latter in excess),
while the LDA-LO result is about 2\%  below the experimental value. As for beyond-local functionals, HSE sllightly underestimates  
PBE results and as a consequence is quite close to experiment, while PSIC and ASIC underestimate their respective local-functional (LDA-PW and 
LDA-LO) references by $\sim$1{\%} on average (a tendency also found\cite{vpsic} in other classes of oxides such as titanates and manganites). 

\begin{figure*}[ht]
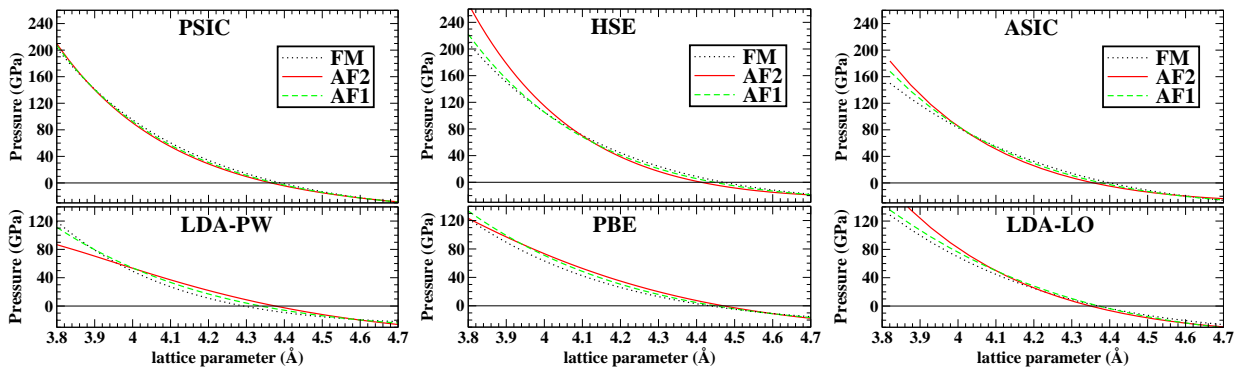

\centering
\includegraphics[clip=, width=0.3\linewidth]{2a}
\includegraphics[clip=, width=0.3\linewidth]{2b}
\includegraphics[clip=, width=0.3\linewidth]{2c}
\caption{(Color on-line) Calculated pressure for MnO in the three considered magnetic phases (see text). Each panel reports results obtained by a 
different energy functional or methodology considered in this work.
\label{mno_press}}
\end{figure*}

\begin{figure*}[ht]
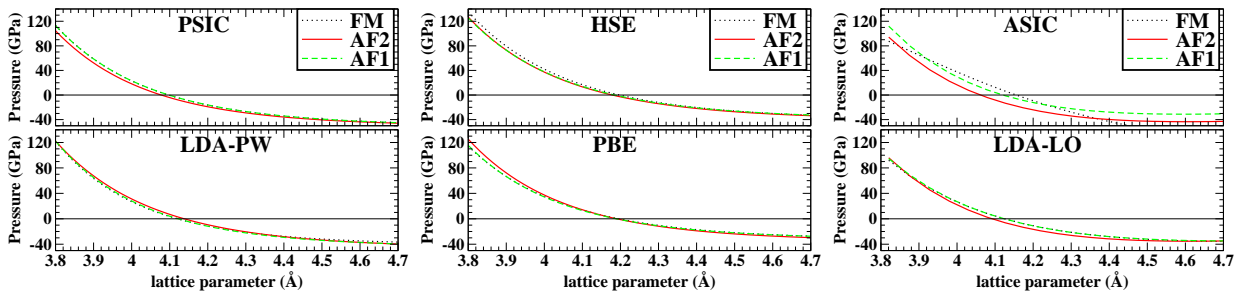

\centering
\includegraphics[clip=, width=0.3\linewidth]{3a}
\includegraphics[clip=, width=0.3\linewidth]{3b}
\includegraphics[clip=, width=0.3\linewidth]{3c}
\caption{(Color on-line) Calculated pressure for NiO in the three considered magnetic phases (see text).
Each panel reports results obtained by a different energy functional or methodology considered in this work.
\label{nio_press}}
\end{figure*}

In Figs. \ref{mno_press} and \ref{nio_press} the calculated pressures for, respectively,  MnO and NiO are reported as a 
function of lattice parameters for the three magnetic phases. The behavior in the region around the equilibrium lattice
constant is expressed by the calculated bulk modulus (B$_0$) in Tab.\ref{tab_lattice}. The advanced functionals 
coherently give an increase of B$_0$ by $\sim$20-30 GPa ($\sim$15\%) with respect to their respective local functionals.
This increase is a consequence of the enhanced 3d state localization and concomitant increase in Coulomb
repulsion under compression which is expected from beyond-local approaches. Concerning the agreement with experiment, 
both PBE and HSE give values within the reported experimental uncertainty. In contrast, LDA-PW gives 
B$_0$ at the higher end of the experimental error bar, thus that the 15\% further increase caused by PSIC pushes the
value of B$_0$ $\sim$40-50 GPa above the experiment. Hence the discrepancy should be seen as due to the LDA-PW performance
(and to the characteristics of the used pseudopotentials), rather than as a failure of the PSIC method in itself. 
This trend can also be observed for the local orbital basis, where the LDA results for B$_0$ are already slightly above the the 
experimental values, and the addition of the ASIC corrections simply pushes the bulk modulus further away.
 
A very interesting feature which emerges consistently from all the methods is the quite similar pressure dependence for different magnetic orderings, 
especially evident for NiO. This looks surprising at first glance, especially at strong compressive pressure, where (as discussed later on) changes 
in the magnetic ordering are related to metal-insulating transitions and to radical changes in the electronic properties. The explanation  is that $e_g$ 
electrons (strongly hybridized with O $p$ states) govern the electronic and magnetic properties, but have only a minor effect on the response to applied 
hydrostatic pressure. In MnO, where  $t_{2g}$ states are also magnetically active, pressure is slightly more sensitive to the specific magnetic ordering
(in Fig.\ref{mno_press} AF$_2$ tends to differ from AF$_1$ and FM, which almost overlap  each other). It should also be noted that in MnO for very 
contracted lattice constants the advanced functionals give pressures a factor of 1.5-2 larger than those of the corresponding local functionals, depending
on the method and specific magnetic phase. At variance, for NiO advanced and local functionals give pressures in the same range. This reflects the larger 
effect of the advanced functionals on the half-filled $t_{2g}$ shell of MnO, which is pushed down in energy and increase its spatial localization and its 
Coulomb repulsion under compression, than on the filled $t_{2g}$ shell of NiO.

\subsection{Magnetic properties upon applied pressure}
\label{res_press}
\subsubsection{Magnetic phase diagram under pressure} 

\begin{figure*}[ht]
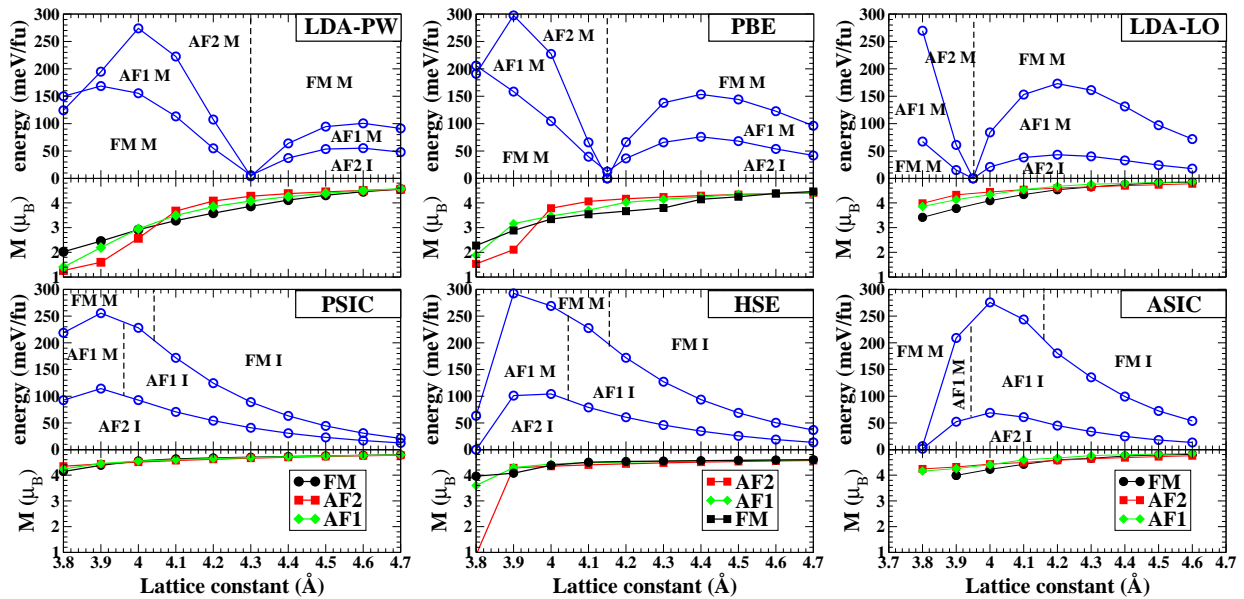

 \centering
\includegraphics[width=0.3\linewidth]{4a}
\includegraphics[width=0.3\linewidth]{4b}
\includegraphics[width=0.3\linewidth]{4c}
 \caption{(Color on-line) Total energies (relative to the ground state) and magnetic moments of FM, AF$_1$ and AF$_2$ 
phases of MnO as a function of lattice parameter, calculated by LDA-PW and PSIC (left panels), PBE and HSE (center panels), 
LDA-LO and ASIC (right panels). The insulating (I) or metallic (M) character of each magnetic phase is also indicated.
Vertical dashed lines indicate phase transitions.       
 \label{mno_energy}}
\end{figure*}

\begin{figure*}[ht]
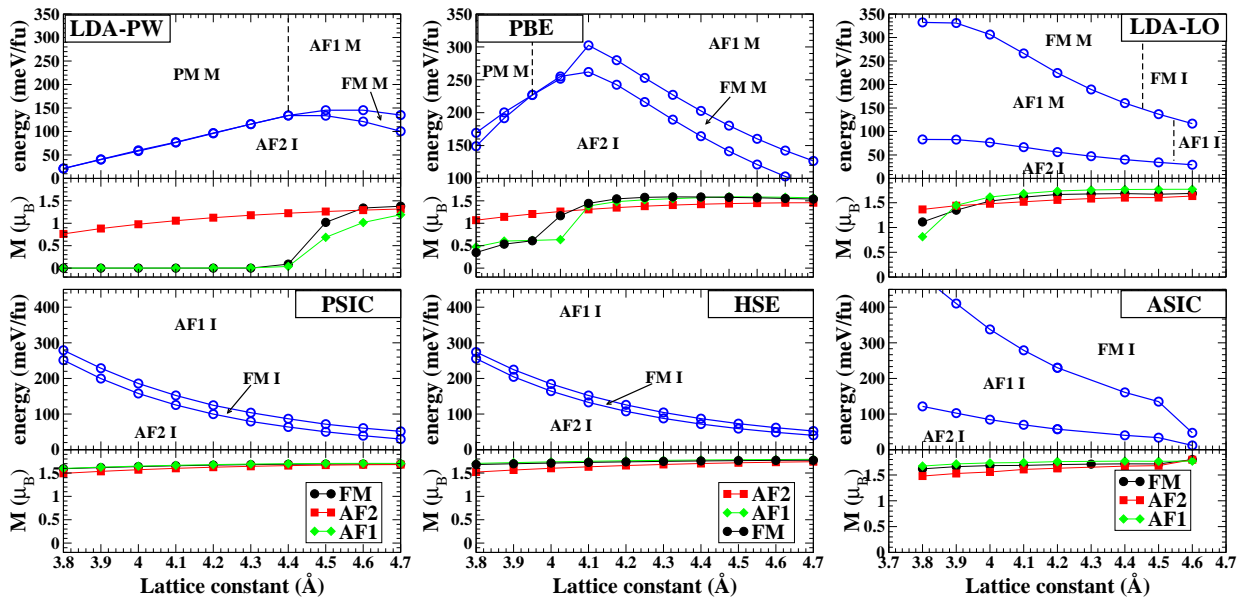

 \centering
\includegraphics[width=0.3\linewidth]{5a}
\includegraphics[width=0.3\linewidth]{5b}
\includegraphics[width=0.3\linewidth]{5c}
 \caption{(Color on-line) Total energies (relative to the ground state) and magnetic moments of FM, AF$_1$ and AF$_2$ phases of NiO as a function of 
lattice parameter, as calculated for the different methods used in this work: LDA-PW and PSIC calculations (left), PBE and HSE calculations (center), 
LDA-LO and ASIC calculations (right). PM indicates Pauli Paramagnetic ordering, I and M insulating and metallic character, respectively. 
Vertical dashed lines indicate phase transitions.       
\label{nio_energy}}
\end{figure*}

Figs. \ref{mno_energy} and \ref{nio_energy} summarize our findings concerning phase stability and magnetic moments for MnO 
and NiO. Each panel reports results obtained by a given energy functional for relative magnetic energies (with respect to 
the most stable magnetic ordering) and their corresponding magnetic moments, as a function of lattice constant. 
Column-wise, panels are ordered according to the code used:  LSDA-PW and PSIC results (left, PWSIC);  
 PBE and HSE  results (center, VASP); LSDA-LO and ASIC results (right, SIESTA).
 	
We start our analysis from MnO results given by LDA-PW, LDA-LO, and PBE (top panels of Fig.\ref{mno_energy}). We can 
capture immediately the substantial similarity of the three approaches: at large volumes all have a region where AF$_2$ is the most 
stable phase and insulating, a metallic AF$_1$ region $\sim$50-100 meV above AF$_2$, and a metallic FM region $\sim$100-150 meV above AF$_2$. 
Going from large- to small-lattice region, all three methods reports a phase transition (indicated by the vertical dashed lines)
through which the AF$_2$ phase yields to a metallic FM region as ground state, which wins over higher-energy AF$_1$ and 
(further above) AF$_2$ metallic phases. The critical value of lattice parameter corresponding 
to the  transition differs in the three approaches  4.3 \AA $\>$ in LDA-PW, 4.15 \AA $\>$ according to PBE, and 3.9 \AA $\>$ for LDA-LO. Looking 
at the corresponding magnetic moments, LDA-PW and PBE show a similarly gradual moment decay when going from 4.7 \AA$\>$ 
up to a threshold of 4.0 \AA $\>$ (LDA-PW) and 4.1 \AA $\>$ (PBE), corresponding, in the pressure scale, to$\sim$40 GPa 
and 80 GPa, respectively. Below this  threshold, magnetic moments fall quite abruptly down to $\sim$ 2 $\mu_B$ 
at 3.8 \AA$\>$ for the stable FM metallic phase (notice that this unphysical collapse described by LDA-PW and PBE is not 
related to the true moment collapse\cite{cohen} in MnO at much higher pressures). At variance, moments calculated by LDA-LO decay gradually 
through the whole lattice constant range, without collapsing as in LDA-PW and PBE. This difference reflects the persistence of the AF$_2$ stability
in a larger parameter range described by LDA-LO.

Now we  analyze the results obtained within beyond-local functionals. Overall, the picture is radically different: for the whole range of lattice 
parameters, the AF$_2$ insulating phase is robustly the ground state, and the spurious phase transition discussed above is absent. Furthermore, all 
three approaches coherently report a stability enhancement (i.e. a roughly linear energy gain) of AF$_2$ for decreasing lattice constant. This effect
is indeed expected as a consequence of the increased Mn $d$-O $p$ covalency and the related strengthening of AF superexchange coupling. The AF$_2$ 
maximum stability is reached at $\sim$3.9 \AA $\>$ according to PSIC (corresponding to an applied compression of $\sim$130 GPa), at $\sim$4.0 \AA $\>$ 
for ASIC (P$\sim$130 GPa) and HSE (P$\sim$90 GPa). The peak of AF$_2$ energy gain with respect to the equilibrium structure is nearly 100\%, from 
$\sim$50 meV/f.u. to more than 100 meV/f.u. according to HSE and PSIC. Above AF$_2$ all our advanced functionals favor the AF$_1$ phase, that, at 
variance with the always insulating AF$_2$ phase, undergoes a metal-insulating transition at 3.97 \AA $\>$ (PSIC), 3.95 \AA$\>$ (ASIC) or 4.05 \AA $\>$
(HSE). Above AF$_1$ resides a FM region, again separated in a large-volume insulating and small-volume metallic sides, with an insulating-metal transition
threshold of 4.07 \AA $\>$ for PSIC, and 4.15 \AA $\>$ for both ASIC and HSE. This consistency  is also reflected in similar values of magnetic 
moments through the whole lattice constant range: all methods describe a moderate decline from $\sim$ 4.7 $\mu_B$ at 4.7 \AA $\>$ to $\sim$ 4.0-4.2 $\mu_B$
(depending on the specific magnetic phase) at $a$=3.8 \AA.    

Interestingly, the remarkably coherent picture delivered by our advanced functionals for MnO is not limited to the predict the same ground-state, 
but also involves ordering and energy differences among the three magnetic phases. This is instrumental to coherently describe finite-temperature 
properties as well, as shown in Section \ref{res_crit}.

Now we move to the analysis of NiO results, summarized in Fig. \ref{nio_energy}, starting again from the phase stability diagram drawn
by the three local functionals (upper panels). At variance with what seen for MnO, now all the methods give the insulating
AF$_2$ phase as stable at any lattice constant; however the competition with the other two orderings is described 
differently: according to LDA-PW, moving from large to small lattice constants there is first a tiny region where the 
AF$_2$ stability increases, reaches maximum at 4.4 \AA$\>$ (thus much above the equilibrium value) and then start falling 
linearly all the way down to 3.8 \AA$\>$. Furthermore, above AF$_2$ the LDA-PW predicts a coexistence of degenerate AF$_1$ 
and FM metallic phases. This scenario can be rationalized looking at the magnetic moments for AF$_1$ and FM 
calculated in LDA-PW: starting from the large lattice constant value of $\sim$ 1 $\mu_B$, the magnetic moment falls rapidly and 
vanishes altogether just at 4.4 \AA$\>$ (i.e. still above the equilibrium value of 4.35 \AA). Below this threshold the 
LDA-PW describes a metallic, Pauli paramagnetic region. On the other hand, the expected Mott-insulating behavior is only 
maintained in the AF$_2$ phase. This feature represents a major shortcoming which seriously hamper the NiO description by LDA-PW. 

PBE shows a similar, although slightly less dramatic failure, since the moment collapse starts to show up for smaller 
lattice constant values (4.0 \AA$\>$ for FM and 4.1 \AA$\>$ for AF$_1$, thus definitely below the equilibrium 4.19 \AA$\>$), 
and the magnetic moment is severely reduced to about 0.5 $\mu_B$, without vanishing completely. Quite distinct is the behavior of LSDA-LO, 
which shows two well energy-separated phases, AF$_1$ above AF$_2$, and then FM further above, both undergoing metal-insulating transitions 
when moving from smaller to larger lattice parameters. This peculiarity is clearly reflected in the calculated magnetic moment: there is no 
collapse at least until 4.0 \AA$\>$, then a rapid downfall starts, but the moments remain quite sizable up to 4.0 \AA$\>$, at variance with 
what seen for PBE and LSDA-PW, and consistently with what is found for MnO as well. This signals the tendency of LDA-LO to conserve sizable
Mott gap and magnetic moments in a range of lattice parameters where plane-wave based implementations describe the system
as already collapsing to metallic paramagnets (as explicitly shown later on in the DOS analysis).         

Moving to the analysis of beyond-LDA functionals, all of them consistently predict the large enhancement of 
AF$_2$ stability upon lattice constant decrease in a wide range around the equilibrium value. 
The artificial moment collapse described by the local functionals is absent, and all the magnetic phases remain 
insulating through the whole lattice constant range. All methods find magnetic moments of about 1.7-1.8 $\mu_B$ 
at large lattice constants, and a very moderate decrease to $\sim$1.5 $\mu_B$ at the smallest lattice constant considered (3.8 \AA).   
Particularly striking is the agreement between HSE and PSIC, both describing a tiny FM region intermediate between AF$_2$ 
and AF$_1$, and a parabolic rise of the AF$_2$ energy gain from $\sim$50 meV at 4.7 \AA$\>$ up to $\sim$250 meV at 
3.8 \AA. On the other hand, ASIC gives a different phase ordering (AF$_2$-AF$_1$-FM), with a smaller increase of the 
AF$_2$ relative stability from $\sim$0 meV at 4.7 \AA$\>$ to $\sim$120 meV at a=3.8 \AA. 

\begin{figure*}[ht]
\centering
\includegraphics[clip=, width=0.45\linewidth]{6a.eps}
\includegraphics[clip=, width=0.45\linewidth]{6b.eps}
\includegraphics[clip=, width=0.45\linewidth]{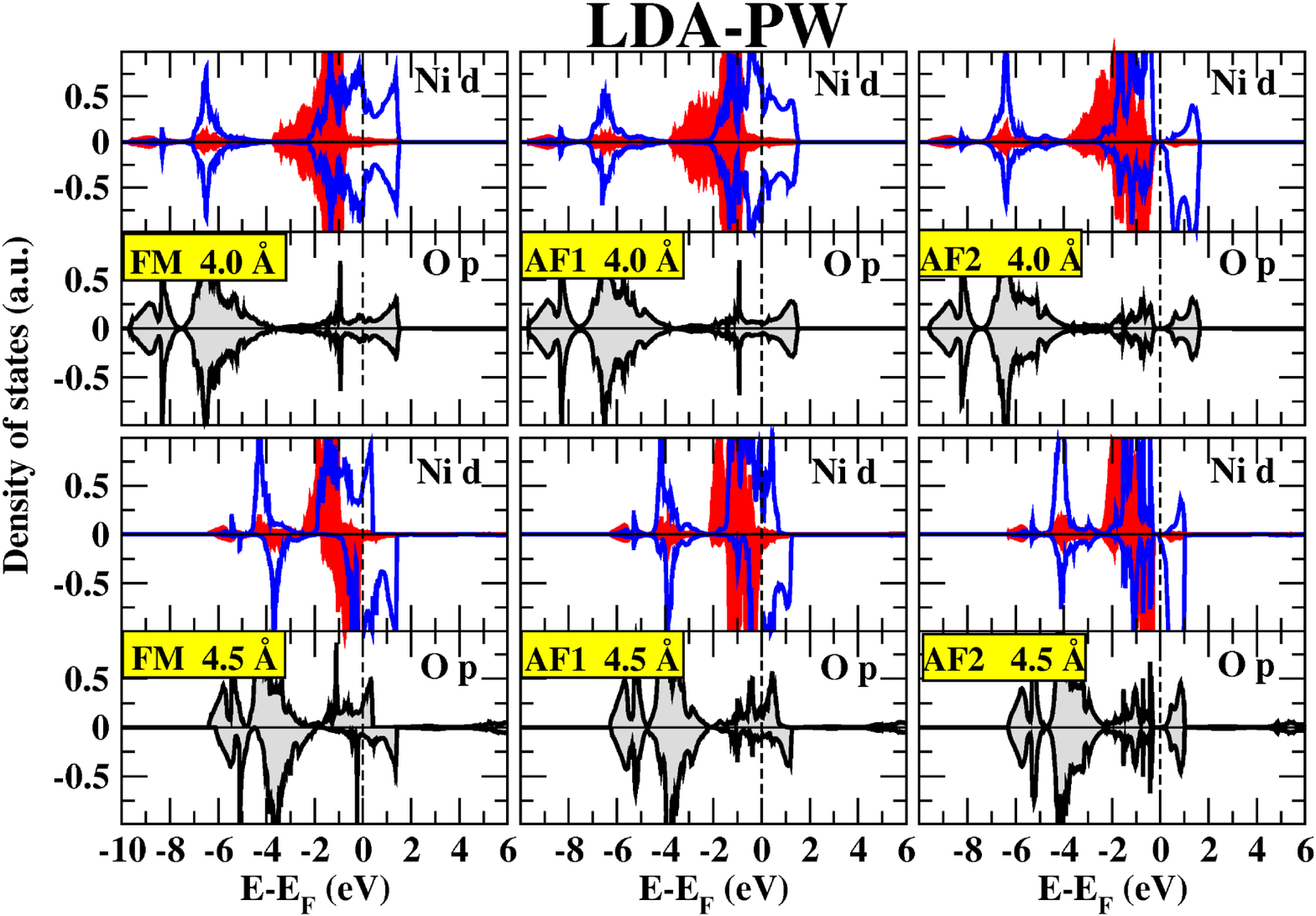}
\includegraphics[clip=, width=0.45\linewidth]{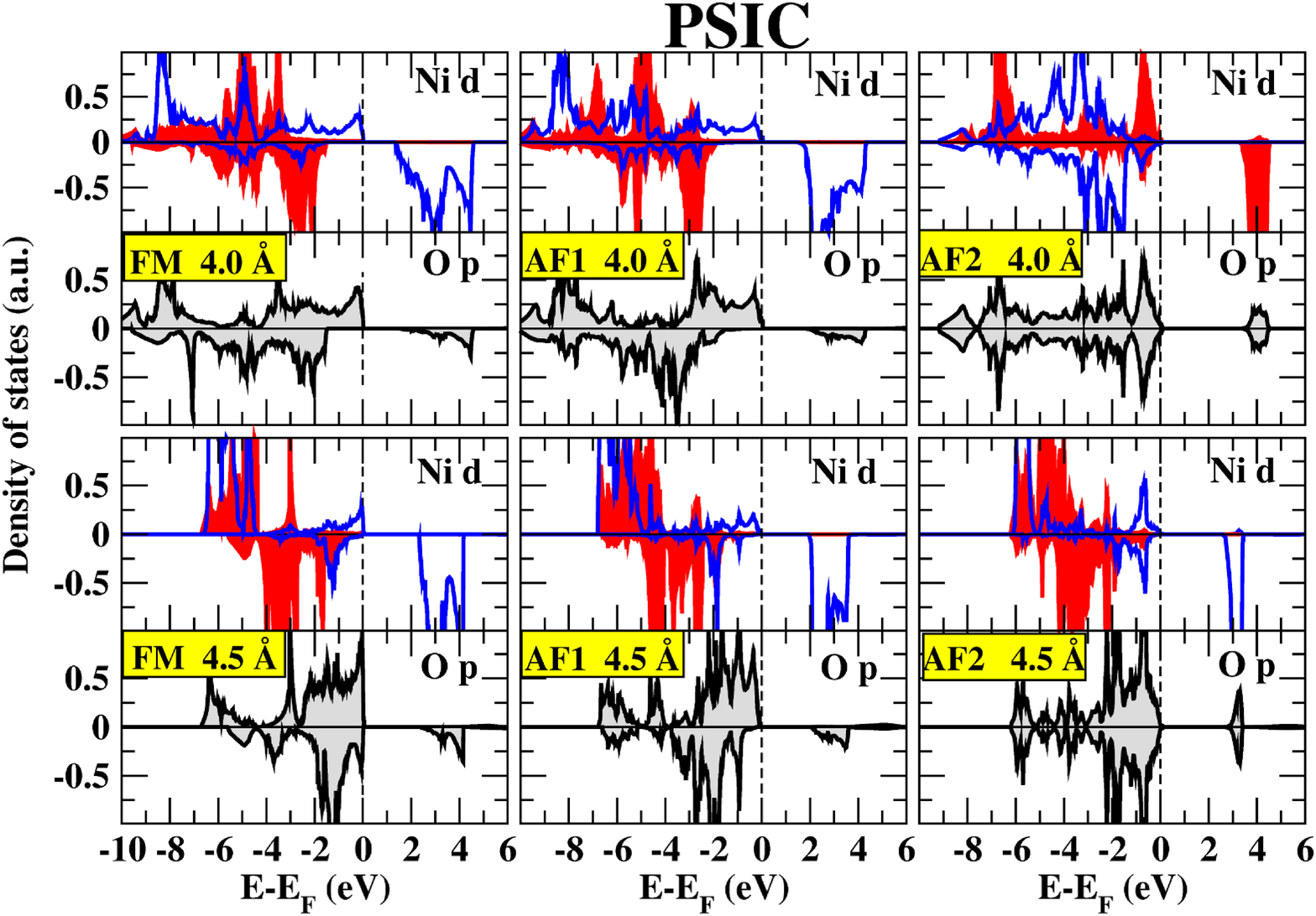}
\includegraphics[clip=, width=0.45\linewidth]{6e.eps}
\includegraphics[clip=, width=0.45\linewidth]{6f.eps}
\caption{(Color on-line) Orbital-resolved DOS for NiO calculated at two lattice constant values (4.0 \AA, and 4.5 \AA) with all our
employed functionals: LDA-LO (top left), ASIC (top right), LDA-PW (middle left), PSIC (middle right), PBE (bottom left)
and HSE (bottom right). Only the DOS for the relevant orbitals are shown: O p (filled gray curves) and Ni d, separated in t$_{2g}$ 
(filled red curves) and e$_g$ (solid blue lines) contributions. Positive and negative curves represent majority and minority contributions. 
\label{nio_dos}}
\end{figure*}

In order to clarify the difference in the magnetic moments under pressure obtained by the different methods, 
we have calculated the orbital-resolved density of states (DOS) for NiO at two representative lattice constants, $a$=4.5 \AA$\>$ 
and $a$=4.0 \AA$\>$, corresponding to situations of tensile and compressive strain. The results are shown in  Fig. \ref{nio_dos}.

Starting from the local-functional results at 4.0 \AA, we notice a radical difference between the two LDA implementations. LDA-LO predicts 
a metallic ground state for both the FM and AF$_1$ phases, with sizeable $e_g$ spin-splitting and substantial magnetic moments on Ni, whereas 
the AF$_2$ phase is found insulating, with a $\sim$1 eV energy gap. On the other hand according to LDA-PW FM and AF$_1$ phases are 
Pauli-paramagnetic, with perfectly compensated spin-densities, while in the AF$_2$ phase a barely visible gap opens up within the $e_g$ manifold. 
PBE results stand somewhat in the middle of the previous two descriptions: for FM and AF$_1$ some Pauli magnetization shows up in the upper
$e_g$ manifold, while for AF$_2$ the magnetic moment is formed, and a Mott gap is about to open. Even at $a$=4.5 \AA$\>$ some differences 
among the local functionals are detectable: according to LDA-LO a minimal gap is opened for each phase, while for LDA-PW only the AF$_2$ is 
insulating. PBE provides well formed magnetic moments in each phase but only AF$_2$ is clearly insulating. 

Going to the results for beyond-local functionals, the expected picture of wide-gap intermediate charge-transfer/Mott insulator described by
the experiments\cite{schuler, shukla, kunes2} is restored. Now the DOS is that of a robust insulator under both lattice expansion and compression, 
with a valence band top populated by a mixture of O p and Ni d states, and the conduction bottom with a majority of e$_g$ and a minor fraction of O p 
states. The energy gap for the AF$_2$ phase ($\sim$3 eV for ASIC and $\sim$3.5 eV for both PSIC and HSE) is in good agreement with the experimental 
value, and even FM and AF$_1$ phases show sizeable gaps of about 1-2 eV. Overall, the similarity of spectral redistributions for $e_g$, $t_{2g}$, and 
O $p$ states (especially for HSE and PSIC) is remarkable.       

In summary, for both MnO and NiO beyond-local functionals deliver a very coherent description of relative phase 
stabilities in the whole examined range of lattice parameters, and predict a clear enhancement of the AF$_2$ phase relative
stability (not described by local functionals) within a wide lattice parameter interval, which suggests the possibility 
of an enhancement of the magnetic ordering N\'eel temperature (T$_N$) upon  applying compressive stress. Before exploring the validity 
of this expectation we will first discuss the evolution of the magnetic coupling constants upon compression. 

\subsubsection{Exchange interactions under pressure}
\label{res_exch}

A few qualitative considerations on magnetic interactions can help to correctly interpret our results. Our calculations find the T=0 
magnetic ground state to be the observed AF$_2$ for both MnO and NiO; however, the detailed magnetic interactions suggest two different scenarios. 
Magnetic coupling between Mn$^{2+}$ 3$d^5$ ions is mediated by half-filled orbitals (thus J$_1$ mainly by $t_{2g}$-$t_{2g}$ and J$_2$ by 
$e_g$-$e_g$ couplings), which are both robustly AF oriented according to superexchange theory.\cite{anderson59,GoodenoughKanamori} Thus we expect J$_1$
and J$_2$ to be both sizable and negative (i.e. AF in our present convention). For Ni$^{2+}$ $d^8$ ions, on the other hand, only $e_g$-$e_g$ coupling is 
magnetically active. Hence we expect a large and negative J$_2$ due to the dominance of covalent superexchange, very small and positive J$_1$ due to 
superexchange-mediated 90$^{\circ}$-oriented $e_g$-(O $p_x$,$p_y$)-$e_g$ orbital coupling, and therefore huge J$_2$/J$_1$ values.

\begin{figure}
\centerline{\includegraphics[clip,width=8.5cm]{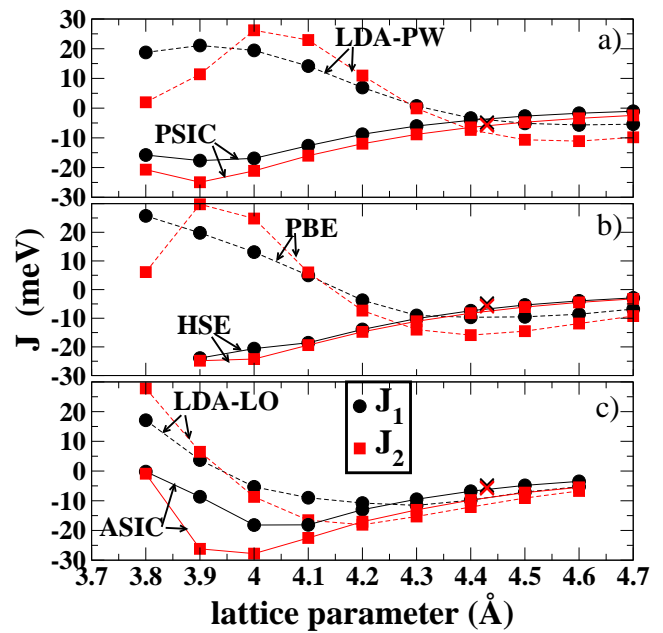}}
\caption{(Color on-line) Exchange-interaction parameters J$_1$ (black filled circles) and J$_2$ (red filled squares) as a function of lattice parameter 
calculated for MnO with various approaches. From top to bottom: a) with PWSIC code by LDA-PW and PSIC functionals; b) with VASP code
using PBE and HSE functionals; c) with SIESTA code using LDA-LO and ASIC functionals. Dashed lines refer to local density functional (LDA-PW, PBE, LDA-LO)
calculations, solid lines represent beyond-local density functionals calculations (PSIC, HSE, and ASIC). Black and red crosses show experimental values for
J$_1$ and J$_2$, respectively, reported for MnO in Table \ref{tab_j}. 
\label{exc-int_mno}}
\end{figure}

These expectations are largely confirmed by our results: in MnO (Fig.\ref{exc-int_mno}) J$_1$ and J$_2$ roughly track each other as function of the 
lattice parameter irrespective of the calculation method. However, it is of the utmost importance to observe the dramatic difference between the 
description of local and advanced  functionals: looking at LDA-PW results [panel a)] J$_1$ and J$_2$ are moderately negative at expanded lattice, then 
upon lattice shrinking they both change sign and grow up to a maximum value at $\sim$ 4 \AA, and finally fall back down as the lattice shrinks further. 
The PSIC almost completely reverses this behavior: J$_1$ and J$_2$ nearly vanish at 4.7 \AA$\>$ (signaling a shorter interaction range with respect to 
LDA), and then grow steadily (in absolute value) on the negative side as the lattice squeezes up. Curiously, LDA-PW and PSIC curves intersect each other
at $\sim$ 4.4 \AA, but this agreement near the equilibrium lattice is just a fortuitous crossing of two otherwise radically different behaviours. 
Notice finally that while in PSIC J$_2$/J$_1$$>$1 at any lattice parameter, in LDA-PW the exchange interaction ratio fluctuates as function of the lattice
constant.

The considerations exposed for LDA-PW and PSIC can be identically repeated for PBE and HSE, respectively (the 
similarity of curves is apparent comparing panel a) and b) in Fig.\ref{exc-int_mno}). The behavior of LDA-LO [panel c)] is more peculiar: 
as previously noticed, the local-orbital implementation seems to cure, though only partially and perhaps accidentally, already
at the LDA level some of the deficiencies of the local functionals related to the self-interaction problem. 
It turns out indeed that the exchange interactions determined by LDA-LO show characteristics similar to those depicted 
by the non-local functionals: at large lattice parameter both J$_1$ and J$_2$ are vanishing and negative, then they 
correctly grow negative upon lattice shortening up to 4.2 \AA  (thus well below the equilibrium structure), but then they
revert slope and turn positive below 4.0 \AA. The ASIC, on the other hand, restores through the whole lattice
range the correct description already seen for PSIC and HSE.

\begin{figure}
\centerline{\includegraphics[clip,width=8.5cm]{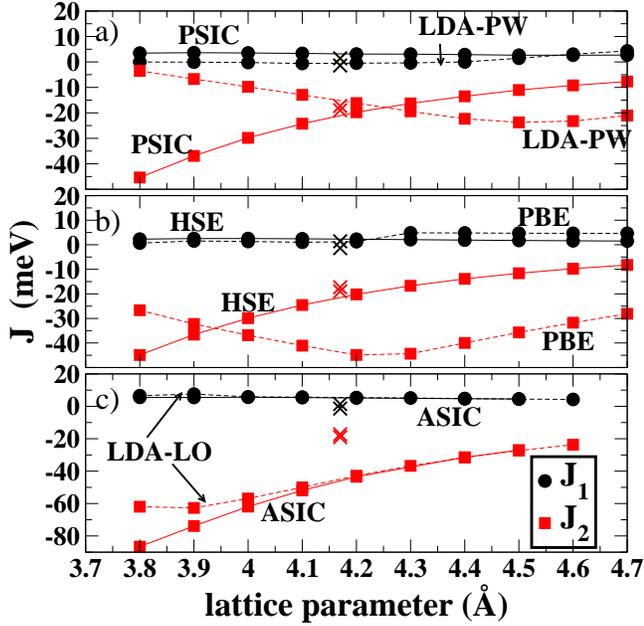}}
\caption{(Color on-line) Exchange-interaction parameters J$_1$ (black filled circles) and J$_2$ (red filled squares) as a function of lattice 
parameter calculated for NiO with various approaches. From top to bottom: a) with PWSIC code by LDA-PW and PSIC functionals; b) with VASP code
using PBE and HSE functionals; c) with SIESTA code using LDA-LO and ASIC functionals. Dashed lines refer to local density functionals calculations 
(LDA-PW, PBE, LDA-LO), whereas solid lines represents beyond-local density functionals calculations (PSIC, HSE, and ASIC). Black and red crosses 
show experimental values for J$_1$ and J$_2$, respectively, reported for NiO in Table \ref{tab_j}. 
\label{exc-int_nio}}
\end{figure}

Now we move to examine NiO (Fig.\ref{exc-int_nio}). As expected, the relative weight of J$_1$ and J$_2$ is very different: all functionals
(both local and beyond-local) describe J$_1$ as very small and positive at any lattice value (black symbols).
On the other hand, the large and negative J$_2$ is again differently described by the two sets of functionals. For 
LDA-PW/PBE and PSIC/HSE (Fig.\ref{exc-int_nio}) we can repeat most of the considerations made for J$_2$ in MnO: the behavior 
with lattice parameter is roughly inverted, although the two curves crosses at different values of the lattice constant (4.2 \AA\, for LDA/PSIC,  
3.9 \AA\, for PBE/HSE). LDA-LO is again a case in its own, as it appears to provide results not far from the ASIC ones in the lattice region above 
4.2 \AA. ASIC qualitative behavior for both J$_1$ and J$_2$ is in line with that of PSIC and HSE, although its J$_2$ appears amplified by nearly a 
factor 2. In consideration of what we have seen so far, the reason of this difference seems to reside more in the methodology (i.e. technical 
implementation: plane wave versus atomic orbitals) rather than in the employed functional. 

From a phenomenological viewpoint, it is important to note the appreciable growth of the exchange interactions for  decreasing lattice parameter in a wide
region extending well below the equilibrium value (i.e. for large applied compression), coherently described by all our beyond-local functionals. 
In the next section we will illustrate the remarkable reverberations of  this behavior on the magnetic ordering temperature. For now, we remark that this
is the expected behavior of the so called covalent exchange, i.e. the shorter the TM-O distance, the stronger the energetic advantage for the O 2$p$ 
ligand states to overlap with the adjacent TM 3$d$ states with unlike spins. This advantage reaches a maximum at a certain compression, after which the
J's start falling. This is the point when the pressure is so strong that minority DOS begins to be appreciably populated and in turn magnetic moments 
start falling.

Overall, the comparative analysis of the three beyond-local functionals is very satisfying, as they furnish a qualitatively and quantitatively coherent
and physically sound description of exchange interactions under pressure for MnO and NiO. There is a large body of data in literature, both theoretical 
and experimental, with which to compare our data, at least at equilibrium. In Table \ref{tab_j} we report our values for the theoretical equilibrium 
structure in comparison with the experimental values and other theoretical predictions obtained by several density functional based approaches 
(see Section I; we neglect tight-binding or shell-model results, which rely on experimental fitting). 

The J's  reported in the Table are determined by finite energy differences, unless otherwise specified. In some cases, the magnetic force theorem 
(MFT)\cite{liechtenstein} based on the exchange-correlation density-functional gauge invariance under infinitesimal spin rotations\cite{solovyev,vignale}
was used. In Ref.\onlinecite{kotani}, exchange interactions and the whole spin-wave spectrum was determined from the poles of spin 
susceptibility.\cite{savrasov,karlsson} As for experiments, results from both inelastic neutron scattering (INS)\cite{kohgi, hutchings} and thermodynamic 
data (TD)\cite{lines, shanker} are reported.

Comparison of J values given by different approaches and considerations on the level of  agreement with the available experimental data must be taken 
very carefully. Differences of the order of a few meV may derive from technical implementation aspects rather than from the basic theory itself, as 
testified by the sizable difference of results obtained by the same theory in different calculations (e.g. our LDA-LO and LDA-PW). 
Concerning MnO, the comparison is further complicated by the close similarity of the two J values. It is expected that local functionals should 
overestimate the J's, due to the underestimation of intra-atomic exchange splitting and the related overestimation of $p$-$d$ hybridization. 
This is indeed verified with LDA-LO and PBE. Notice, however, the relevant exception of LDA-PW (overall, the least accurate of the three local 
functionals) which deliver J's in excellent agreement with the experiment: this fortuitous agreement was previously explained as consequence of the
unphysical AF$_2$ to FM transition occurring in LDA-PW near the equilibrium structure. On the other hand, our beyond-local functionals perform quite 
satisfactorily, apparently ranking among the closest to experiment both in absolute terms and concerning the J$_2$/J$_1$ ratio (1.16 for INS data, 
1.1 for HSE, 1.4 and 1.5 according to ASIC and PSIC). 

For NiO the analysis is simpler, as J$_1$ is very small and a qualitative comparison can be done on the base of J$_2$ only.
We have already commented that the LDA-PW is grossly inadequate, thus it can be left aside. As for LDA-LO and PBE, they 
deliver sizably overestimated J$_2$. As for beyond-local functionals, we have previously seen that that J's calculated 
by PSIC and HSE almost overlap each other throughout the lattice parameter range. The agreement with values drawn from 
neutron experiments\cite{hutchings} is indeed quite satisfying. Nevertheless, the slight PSIC underestimation of the 
equilibrium lattice parameter (4.09 \AA $\>$ against the near experimental-matched 4.18 \AA$\>$ of HSE) reverberates in 
a $\sim$15\% overestimation of J$_2$. On the other hand, ASIC delivers J's that barely differ from the corresponding 
LDA-LO values, and are amplified by nearly a factor 2 with respect to PSIC and HSE. Looking at previous literature, we 
found a substantial agreement of PSIC and HSE with GGA+U calculations of Ref.\onlinecite{zhang} (here the J's are also 
calculated as a function of lattice parameter) and with other types of hybrid functionals\cite{moreira} as well. 
On the other hand, both unrestricted HF\cite{moreira}, full SIC in LMTO approach (SIC-LMTO)\cite{kodde} and the local SIC 
(LSIC) (a KKR-based implementation of the self-interaction correction method\cite{huges,dane,fischer}) tend to an excessive
electronic localization, which thus turns into a slight underestimation of the exchange interactions.

\begin{center}
\begin{table}[ht]
\caption{Exchange interaction parameters for MnO and NiO (in meV) calculated in this work, compared to experimental and
theoretical values from previous works.}
\begin{tabular}{lccp{1cm}lcc}
\hline
                        & \multicolumn{2}{c}{MnO}&&          & \multicolumn{2}{c}{NiO}     \\  
                       &   J$_1$    &    J$_2$  &&           &   J$_1$    &    J$_2$       \\
\hline\hline
\multicolumn{7}{c}{Experiment}\\
&&&&&&\\
INS$^a$                & -4.8       &  -5.6     && INS$^b$    &  1.4  &  -19.0 \\
TD$^c$                 & -5.4       &  -5.9     &&  TD$^d$    & -1.4  &  -17.3 \\
\hline\hline     
\multicolumn{7}{c}{This work: local functionals}\\ 
&&&&&&\\
LDA-PW                  &  -2.7      &    -6.3   && LDA-PW    & -0.5  &  -14.7         \\
LDA-LO                  &  -10.6     &   -13.6   && LDA-LO    &  5.4  &  -43.5         \\
PBE                     &  -9.5      &   -14.9   && PBE       &  1.2  &  -44.5         \\
&&&&&&\\
\multicolumn{7}{c}{This work: advanced functionals}\\ 
&&&&&&\\
PSIC                    &  -5.0      &    -7.6   && PSIC      &  3.3  &  -24.7         \\
HSE                     &  -7.0      &    -7.8   && HSE       &  2.3  &  -21.0         \\
ASIC                    &  -8.0      &   -11.3   && ASIC      &  5.2  &  -45.0         \\
\hline\hline  
\multicolumn{7}{c}{Previous calculations}\\ 
&&&&&&\\
LSDA$^e$  (MFT) & -13.2      &    -23.5  &&                     &        &                 \\ 
LDA+U$^e$ (MFT) &  -5.0      &    -13.2  &&GGA+U$^l$            &   1.7  &  -19.1 \\
OEP$^e$   (MFT) &  -5.7      &    -11.0  &&SIC-LMTO$^m$         &   1.8  &  -11.0  \\
PBE+U$^f$       &  -4.4      &    -2.3   &&Fock-35$^n$          &   1.9  &  -18.7  \\
PBE0$^f$        &  -6.2      &    -7.4   &&B3LYP$^n$            &   2.4  &  -26.7  \\
HF$^f$          &  -1.5      &    -2.32  &&UHF$^n$              &   0.8  &  -4.6   \\
QPGW$^g$        &  -2.8      &    -4.7   &&QPGW$^g$             &  -0.8  &  -14.7  \\
LSIC$^h$        &   1.4      &    -3.3   &&LSIC$^h$             &   2.8  &  -13.9 \\
LSIC$^h$ (MFT)  &  -1.8      &    -4.0   &&LSIC$^h$     (MFT)   &   0.3  &  -13.8 \\
B3LYP$^i$       &  -5.3      &    -11.0  &&                     &        &         \\
\hline\hline
\multicolumn{7}{l} { a): Ref.\onlinecite{kohgi}, b): Ref.\onlinecite{hutchings}, c): Ref.\onlinecite{lines} 
                     d): Ref.\onlinecite{shanker} } \\
\multicolumn{7}{l} { e): Ref.\onlinecite{solovyev}, f): Ref.\onlinecite{franchini}, g): Ref.\onlinecite{kotani}, 
                     h): Ref.\onlinecite{fischer} } \\
\multicolumn{7}{l} { i): Ref.\onlinecite{feng},     l): Ref.\onlinecite{zhang},    m): Ref.\onlinecite{kodde}, 
                     h): Ref.\onlinecite{moreira} } \\
\end{tabular}
\label{tab_j}
\end{table}
\end{center}

\subsubsection{Critical transition temperatures under pressure}
\label{res_crit}

\begin{figure}
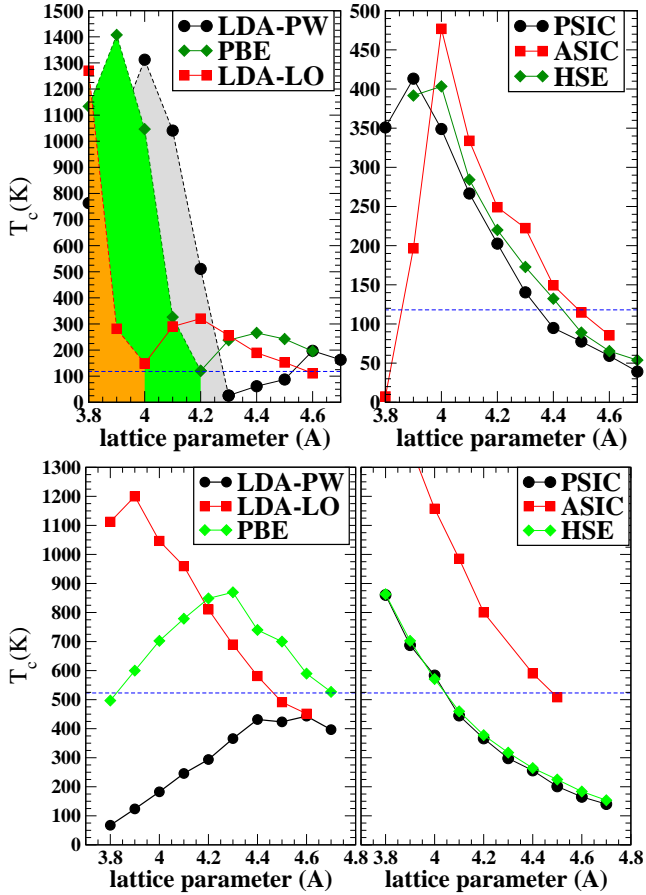

\centerline{\includegraphics[clip,width=8.5cm]{9a.eps}}
\centerline{\includegraphics[clip,width=8.5cm]{9b.eps}}
\caption{(Color on-line) Critical temperatures as a function of lattice constant for MnO (top panel) and NiO (bottom) calculated by 
simulated-annealing MonteCarlo simulation of the Heisenberg Hamiltonian in Eq.\ref{heisen}. Left and right panels separate local and advanced 
functionals used to determine each set of (J$_1$, J$_2$). The shaded areas in the top-left panel indicate FM metallic regions; apart from that, 
each curve separates low-T insulating AF$_2$ from high-T insulating PM regions (PM stands for Pauli Paramagnetic ordering). The dashed horizontal 
lines indicate experimental T$_N$ values.
\label{tc}}
\end{figure}

Fig. \ref{tc} reports critical temperatures for MnO and NiO calculated with the Heisenberg Hamiltonian given 
in Eq.\ref{heisen} and solved through classical MonteCarlo (MC) simulated annealing technique.\cite{mc} Values calculated  
at equilibrium and at experimental volume are reported in Tab. \ref{tab_tcrit}, in comparison with the experiment
and other theoretical predictions. 

We  start discussing the case of MnO as described by our local density functionals. As expected, MC calculations 
describe fictitious phase transitions (consequence of the spurious magnetic moment collapse previously discussed) from AF$_2$ to 
FM metallic magnetic phase (highlighted by the dashed areas) while moving from large to small lattice parameters. 
The onset of this transition depends on the method: it is especially harmful in LDA-PW as it occurs near (just below) the 
theoretical lattice value (4.35 \AA); much less damaging in LDA-LO, where it happens at a much larger lattice contraction (4 \AA);  
PBE is halfway the two previous cases. In all cases this phase transition dramatically alters the critical temperature behavior, 
which is expected to grow (at least in some interval around equilibrium) as the lattice parameter contracts. Furthermore
in all cases the predicted T$_N$ (see Tab. \ref{tab_tcrit}) remains quite distant from the experimental T$_N$=118 K
(horizontal dashed line in the Figure). Notice again that while LDA-LO and PBE overestimates T$_N$ at the equilibrium structure, 
the spurious phase transition in LDA-PW cause T$_N$ to be smaller than the experimental values, and paradoxically not too far 
(at experimental lattice constant) from the experiment. This is another manifestation of the fortuitous agreement already pointed out
in the illustration of the exchange interactions.

Conversely, the advanced functionals deliver for MnO a nicely consistent picture, with T$_N$ growing steadily from very large lattice constants 
(T$_N$ $\rightarrow$0) up to 3.9-4.0 \AA, and peaking at $\sim$410 K (PSIC and HSE) or 480 K (ASIC). The corresponding pressure is around 120$\pm$20 
GPa depending on the approach (see Fig.\ref{mno_press}). There is a minor offset between the three methods, due to the slight increase in J's  
moving from PSIC to HSE to ASIC. However, if calculated at their respective equilibrium values, both PSIC and HSE predict a T$_N$ which is
almost spot-on to the experimental value.     

Moving to the analysis of NiO, the three local functionals behave quite differently from each other, although PBE and 
LDA-LO fortuitously cross nearby the equilibrium value. So, it is all the more remarkable that their beyond-local 
counterparts are capable to rebuild a very consistent picture, with T$_N$ linearly growing along with the decrease of
lattice parameter. The near-overlap of HSE and PSIC is especially striking, and already noticed for the J's. The ASIC curve 
appears translated upward by nearly a factor 2, coherently with the overestimate of the dominant J$_2$ value. Notice that for 
NiO there is no lattice parameter turning point in the considered range, thus T$_N$ keeps growing up to 3.8 \AA, corresponding 
to a pressure of about 100-120 GPa (see Fig.\ref{nio_press}). The agreement with the experiment for NiO is much 
less outstanding than for MnO, as both HSE and PSIC remains below the experimental T$_N$=523 K by $\sim$20\% (at the equilibrium structure
the PSIC value is actually not too far from the experiment thanks to its underestimated lattice parameter), while ASIC is close to the much 
overestimated LDA-LO value. 

The disagreement between {\em ab initio} theoretical estimates and the experimental value of T$_N$ for NiO is typical and well documented in literature
(see e.g. Ref.\onlinecite{fischer} and references therein) and a full discussion on the subject is beyond our present scope. We only remark that the 
mismatch with HSE and PSIC is somewhat puzzling, in consideration of the excellent agreement of the calculated J$_2$ with the experimental value. 
In fact, even using experimental sets of J's, the MC-calculated T$_N$ would change only marginally our theoretical value: as a useful reference we 
also report in the Table the T$_N$ of MnO and NiO calculated by MC using sets of experimental J's reported in Table \ref{tab_j}. It turns out that 
these T$_N$ underestimate by $\sim$ 30\% the directly measured T$_N$ (they are even lower than those obtained with our calculated J's, since the slight
overestimation of our calculated J's helps in shifting up the predicted T$_N$). The discrepancy between experimental J's and T$_N$ somewhat points out 
to possible inadequacies of the employed Heisenberg model, possibly due to further terms (not included in Eq.\ref{heisen}) which might be important at the 
relatively high ordering temperature of NiO. 

In Tab.\ref{tab_tcrit} we compare our results with some previous theoretical predictions (we omit the many mean-field approximations which are 
known \cite{anderson59} to grossly overestimate the critical temperature). Ref. \onlinecite{fischer} proposes T$_N$ obtained by disordered local moments
(DLM), random-phase approximations (RPA) and MC, based on the MFT-calculated J's given in Table \ref{tab_j}. The DLM values are closest to our HSE and 
PSIC estimate, while RPA and MC values for NiO are larger than our values despite smaller J's, due to the debatable inclusion in Ref.\cite{fischer} of 
the quantum rescaling factor (S+1)/S.\cite{wan} Ref.\onlinecite{wan} also proposes MFT-calculated J's, derived from LDA, LDA+U, and LDA 
plus dynamical mean field approach solved through cluster exact diagonalization (CED). The latter seems to restore an outstanding agreement 
with the experimental T$_N$ for NiO\cite{note}.

\begin{center}
\begin{table}[ht]
\caption{Critical temperatures (K) for MnO and NiO calculated in this work at equilibrium and experimental (in brackets) lattice constant, compared 
with experimental (neutron diffraction, ND) and theoretical values from previous works. As a reference we also report MC-calculated values obtained 
by using the experimental J's (see text for discussion).}
%\begin{tabular}{lcc}
%\begin{tabular}{lp{2cm}p{2cm}}
\begin{tabular}{p{2.5cm}p{2cm}p{2cm}}
\hline
                       & MnO         &       NiO     \\  
\hline\hline
\multicolumn{3}{c}{Experiment}\\
&&\\
ND$^{a,b}$              &  118      &       523      \\
Expt. J's               &  85$^c$   &       340$^d$  \\
Expt. J's               &  90$^e$   &       300$^f$  \\  
\hline\hline     
\multicolumn{3}{c}{This work: local functionals}\\
&&\\
LDA-PW                  &   55 (71)  &   272 (280)   \\
LDA-LO                  &  220 (176) &   967 (851)   \\
PBE                     &  249 (257) &   824 (829)   \\
&&\\
\multicolumn{3}{c}{This work: advanced functionals}\\
&&\\
PSIC                    &  116 (89)  &   458 (387)   \\
HSE                     &  125 (116) &   393 (400)   \\
ASIC                    &  182 (137) &  1048 (850)  \\
\hline\hline  
%Previous works          &            &               \\
\multicolumn{3}{c}{Previous works}\\
&&\\
LSIC$^g$ (DLM)&  126       &    336        \\
LSIC$^g$ (RPA)&  87        &    448        \\
LSIC$^g$ (MC) &  90        &    458        \\
LDA$^h$  (MC) &  423       &    965        \\     
LDA+U$^h$(MC) &  240       &    603        \\
CED$^h$  (MC) &  172       &    519        \\
\hline\hline
\multicolumn{3}{l} { a): Ref.\onlinecite{roth}, b): Ref.\onlinecite{shull}, c) Ref.\onlinecite{kohgi}, d) Ref.\onlinecite{hutchings},  
e) Ref.\onlinecite{lines}.}\\
\multicolumn{3}{l} { f) Ref.\onlinecite{shanker}, g) Ref.\onlinecite{fischer}, h) Ref.\onlinecite{wan}. } \\
\end{tabular}
\label{tab_tcrit}
\end{table}
\end{center}

\section{Summary and Conclusions}
\label{concl}

It is fair to affirm that the overall account of the structural, electronic, and magnetic  properties of MnO and NiO provided by the advanced functionals
is overall quite satisfying, internally consistent, and in good agreement with experiments. In particular, HSE shows a remarkable quantitative agreement 
with experiments on most examined properties; the PSIC, perhaps surprisingly when considering the substantially different conception at the basis of 
their theoretical constructions, is quite comparable with HSE results, and in some cases in spectacular quantitative agreement (e.g. the NiO exchange 
interactions vs. lattice constant). An important persistent shortcoming of PSIC or ASIC, however, is the prediction of structure: the predicted lattice 
constant is below experiment by $\sim$1-2\%. This tendency to deliver smaller-than-optimal structural parameters was also encountered in other 
situations\cite{vpsic}, and it is probably not an isolated occurrence, but rather a characteristics of the PSIC/ASIC method. In perspective, we expect that 
this drawback could be overcome by adopting the GGA (e.g. PBE) instead of the LDA as reference functional upon which to build the PSIC projector 
(see Refs.\onlinecite{fs,ff-coll} for details). This would probably lead, as in the case of HSE, to a moderate volume reduction compared to the slightly 
overestimated GGA volume, hence probably to an end product much nearer to experiments.

Some additional considerations concern  the comparison of different  implementations of the same theory, i.e. LDA-PW vs. LDA-LO and PSIC vs. ASIC. 
Interestingly, some visible quantitative differences appear already at level of local functionals. Hence, these are not induced by the ASIC/PSIC formulation, 
which in fact tends to re-equalize the respective descriptions. The amplitude of the PSIC vs. ASIC discrepancy is only minor for MnO, and a bit more 
pronounced for NiO, especially concerning the amplitude of J$_2$. This is not particularly surprising, since different choices of local orbital basis 
set and/or pseudopotentials can easily alter meV-scale quantities. In fact we can find in the literature a wide spread of values for exchange interactions 
predicted by different implementations of the same theory (even at the LDA level). Not by chance the different basis set has more impact on NiO exchange 
interactions than in the MnO ones, since the e$_g$ charge spreads more in space than the highly localized t$_{2g}$ one, and as such it is clearly more sensitive 
to the incomplete description of the intersticial space due to local orbital basis. As a trade off, this larger error bar is compensated by the ASIC 
capability to tackle large-size systems (up to a few thousands atoms) which in fact are impracticable for the, in principle, more accurate 
plane-wave approaches such as PSIC and HSE. A valuable testing ground of this capability may be the field of oxide interfaces, where a plethora of new 
exciting phenomena have been recently discovered. It will be interesting to evaluate whether PSIC/ASIC or HSE are capable to describe e.g. composite
systems such as an interface between a metal and insulator, where the degree of charge localization at the two sides is rather different.       

Finally, our evaluation of the exchange-interaction parameters and of the N\'eel temperatures requires a mention. While the calculated J's of both
MnO and NiO are found in satisfying agreement with the experiments, only for the first an equally satisfying T$_N$ is predicted by the MC-solved Heisenberg 
Hamiltonian. This discrepancy can stimulate debate and more work devoted to investigate the possible inadequacies of the Heisenberg Hamiltonian at high 
temperature, an aspect that has not been sufficiently stressed in previous literature.

In conclusion, we have presented  a systematic analysis of the structural and magnetic properties of MnO and NiO under applied pressure, including 
ground-state and finite temperature properties, by using a range of standard and advanced first-principle approaches. The advanced techniques (HSE, PSIC, 
ASIC) describe very consistenly the behavior of the exchange interactions in a wide range of lattice constant values around the equilibrium structure, 
showing an overall agreement with experiments. This places such methods among the most accurate now available to the first-principles community. 
Our results establish a benchmark of accuracy for innovative techniques aimed at the determination of the magnetic properties of magnetic oxides.

\acknowledgments

Work supported in part by the European Union FP7 program and Department of Science and Technology of India under the joint Eu-India 
project ``ATHENA'' (grant Agreement N. 233553); the Italian Institute of Technology under Project NEWDFESCM; the Italian Ministry of 
University and Research under PRIN Project n. 2008FJZ23S (Project ``2-DEG FOXI''); the Fondazione Banco di Sardegna via a 2008-10 
grant; CASPUR supercomputing center; Consorzio Cybersar. Supercomputer time in Austria was provided by the Vienna Scientific Cluster. 
Pinaki Majumdar acknowledges support through a DAE-SRC Research Investigator Award. C.D.~Pemmaraju acknowledges Science Foundation
of Ireland for financial support (Grant No. 07/IN.1/I945). Computer time in Ireland was provided by the HEA IITAC project managed
by the Trinity Center for High Performance Computing and by ICHEC.


\begin{thebibliography}{10}
\bibitem{note}
A proper comparison with our data cannot be drawn as the employed Heisenberg Hamiltonian 
and the J's are not reported.
\end{thebibliography}
\end{document}